\documentclass[a4paper,keeplastbox]{jacow}
\usepackage{pdfpages,multirow,ragged2e}
\makeatletter\ifboolexpr{bool{xetex}}
{\renewcommand{\Gin@extensions}{.pdf,%
    .png,.jpg,.bmp,.pict,.tif,.psd,.mac,.sga,.tga,.gif,%
    .eps,.ps,%
  }}{}
\makeatother
\ifboolexpr{bool{xetex} or bool{luatex}} 
 {}                                      
 {\usepackage[utf8]{inputenc}}           

\usepackage[USenglish]{babel}
\usepackage{lineno}

\begin{document}
\title{MATLAB simulations of the Helium liquefier\break in the FREIA laboratory}
\author{E. Waagaard, R. Ruber, V. Ziemann, FREIA, Uppsala University}
\maketitle
\begin{abstract}
We describe simulations that track a state vector with pressure, temperature,
and gas flow through the helium liquefier in the FREIA laboratory. Most
components, including three-way heat exchangers, are represented by
matrices that allow us to track the state through the system. The only
non-linear element is the Joule-Thomson valve, which is represented by
a non-linear map for the state variables. Realistic properties for the
enthalpy and other thermodynamic quantities are taken into account with
the help of the CoolProp library. The resulting system of equations is
rapidly solved by iteration and shows good agreement with the observed
LHe yield with and without liquid nitrogen pre-cooling.
\end{abstract}
%
\section{Introduction}
\begin{figure}[htb]
\begin{center}
\includegraphics*[width=0.99\columnwidth]{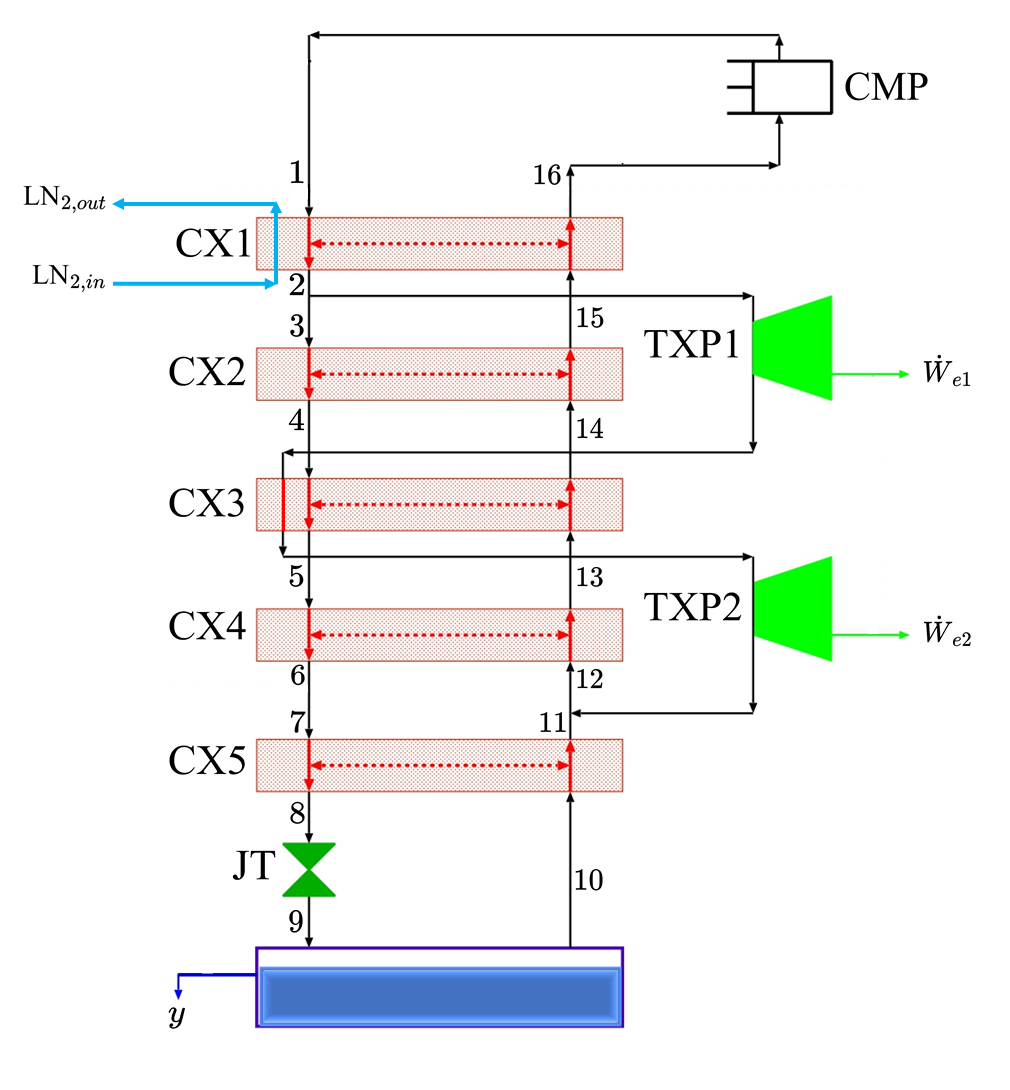}
\end{center}
\caption{\label{fig:FL}Schematics of the FREIA liquefier.}
\end{figure}

The Linde L140 helium liquefier in the FREIA laboratory~\cite{FREIA} at Uppsala
University provides liquid helium 
to cool the superconducting
spoke-cavities for the ESS~\cite{SPOKE} as well as crab cavities~\cite{CC} and
dipoles correctors~\cite{COR} for the HL-LHC upgrade before testing their performance.
The liquefier was delivered without detailed descriptions of its internals, but
we considered this useful anyway, and deduced the schematics, shown in Fig.~\ref{fig:FL},
from the operator interface of its control system and the accompanying documentation.
\par
The warm gas (up to 42\,g/s) leaves the compressor, labeled CMP, at about 13\,bar and
passes through five counterflow heat exchangers (CXn) in the left (warm) branch of the
liquefier, before a Joule-Thomson valve (JT) expands it and cools it to about 4\,K at 1.2\,bar
into the two-phase reservoir, shown at the bottom of Fig.~\ref{fig:FL}, from which the liquid
helium is extracted. The cold gas returns towards the compressor through the cold side of the
five heat exchangers, such that it cools the gas on the warm side. A fraction of the
gas is directed from the warm side at point~2 and passes through two turbo-expanders
(TXPn) that expand the gas and thereby extract energy. The expanded gas after TXP1 passes
near the warm side of CX3, which helps to cool the gas on the warm side, before TXP2
cools it further and returns the cold gas to the cold side at point~11. Finally, liquid
nitrogen can be passed through a third branch of CX1, where it efficiently helps to
cool the gas that arrives in the liquefier with ambient temperature.
\section{Simulation}
We simulate the liquefier with MATLAB~\cite{EW}, where we characterize the state of
the gas at  each point in the liquefier by the temperature $T$, the pressure $P$,
and the gas flow $Q$. All other thermodynamics potentials, such as the enthalpy,
are then calculated with CoolProp~\cite{CP}, which, as a matter of fact, can
convert any two (intensive) thermodynamics quantities to all others, which we
extensively use in our simulation. Moreover, real properties, such as heat capacities
or specific enthalpies of real gases are treated in a consistent way. The action of
CMP, CX, and TXP on the state can be represented by matrices. Only the JT-valve is
a non-linear element, such that we determine the equilibrium condition of the
system by finding a zero of a one-dimensional non-linear function (depending on 
the extracted liquid) that internally solves all the linear relations. This makes
solving the system very fast. But before discussing the full system, we briefly
address the actions of the different elements on the state.
\section{Counterflow heat exchangers}
\label{sec:HX}
\begin{figure}[htb]
\begin{center}
\includegraphics*[width=0.42\columnwidth]{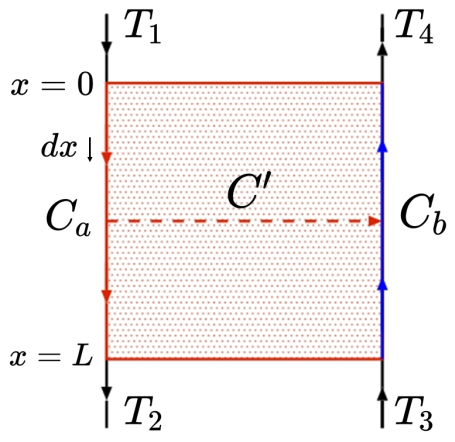}
\includegraphics*[width=0.56\columnwidth]{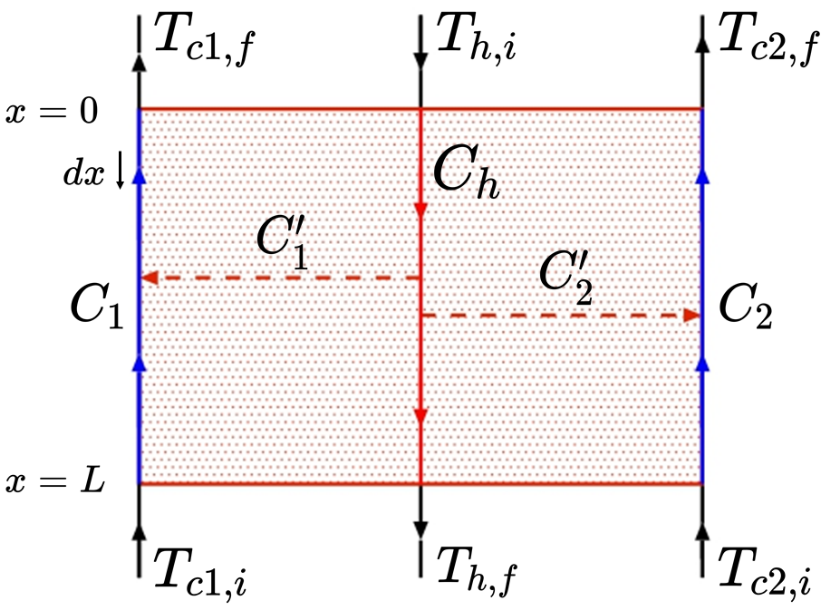}
\end{center}
\caption{\label{fig:CX}Two- (left) and three-way (right) heat exchanger.}
\end{figure}
The left side of Fig.~\ref{fig:CX} shows the two-way heat exchanger that we use for CX2, CX4, and CX5.
The warm gas with temperature $T_1$ enters at the top left and leaves at the bottom left
with $T_2$, whereas the cold gas enters enters from the bottom right with $T_3$ and leaves
the heat exchanger at the top right with~$T_4.$ From the enthalpy balance in a slice $dx$
of the warm and cold gas with heat capacities $C_a$ and $C_b$ and the heat exchanged
with specific heat capacity (per meter) $C'$, we find $d\dot H_a=-C_a dT_a$,
$d\dot H_b=-C_b dT_b$, and $d\dot H=C'(T_a-T_b)dx$, where $\dot H$ is the enthalpy flow
in the respective channels. Combining the equations leads to
\begin{equation}
  \frac{dT_a}{dx}=-\frac{C'}{C_a}(T_a-T_b)
  \quad\mathrm{and}\quad
  \frac{dT_b}{dx}=-\frac{C'}{C_b}(T_a-T_b)\ ,
\end{equation}
which are two coupled linear differential equations that are trivial to solve. Matching
the boundary conditions to $T_1,\dots,T_4$ then leads to the set of linear equations
\begin{equation}
  T_2=T_1-\eta(T_1-T_3)
  \quad\mathrm{and}\quad
  T_4=T_3+\eta\frac{C_a}{C_b}(T_1-T_3)
\end{equation}
where the efficiency $\eta$ is given~\cite{EW} in terms of the heat capacities and the
length $L$ of the heat exchanger. Integrating the enthalpy flow $d\dot H$ along $dx$,
we find~\cite{EW} the heat conduction constant $C_H$, which allows us to calculate
the enthalpy flow from the warm to the cold side as $\Delta H=C_H(T_1-T_3)$. In addition, $C'$ can easily be generalized to three dimensions,  as it is given by the product of the
heat transfer coefficient and the contact area.
\par
We treat the three-way heat exchangers CX1 and CX3 in much the same way, in our case with a hot flow (subscript $h$) and two cold flows (subscript $c$). Using the notation
from the right-hand side in Fig.~\ref{fig:CX}, we find the temperature change from the heat balance in a slice
$dx$ to be 
\begin{align}
&C'_1dT_{c1}/dx=-C'_1(T_h-T_{c1})\\
&C_2dT_{c2}/dx=-C'_2(T_h-T_{c2})\\
&C_hdT_h/dx=-C'_1(T_h-T_{c1})-C'_2(T_h-T_{c2})
\end{align}
which is a linear system that can be
solved analytically. As before, matching the boundary conditions gives us linear
equations that relate the output temperatures $T_{c1,f}, T_{c2,f},$ and $T_{h,f}$ to
the input temperatures  $T_{c1,i}, T_{c2,i},$ and $T_{h,i}$. The details can be found
in~\cite{EW}.
\section{Turbo-expanders}
\label{sec:TX}
A fraction $x$ of the mass flow $Q_2$ at point~2 is directed to the branch with the two
turbo-expanders. We assume that they operate isentropically with $P_{in}/P_{out} =(V_{out}/V_{in})^{\gamma}$,
where $\gamma=5/3$  for a monatomic ideal gas, and that the maximum speed is determined by
the speed of sound at the respective temperatures, as supersonic velocities can reduce the turbo-expander efficiency. Following~\cite{FLYNN}, together with
the ideal gas law, this leads to the linear equations that relate the state variables
of the output to the input variables: $T_{out}=T_{in}/2$ and $P_{out}=P_{in}/5.64.$
\par
The mass flow through the branch with the two expanders is constant and returns to
the cold return line at point~11, where we need to ``mix'' the gas flow $Q_{\textrm{TXP}}$
coming from the expanders and the gas coming from the reservoir. In order to satisfy
the mass flow balance we have $Q_{12}=Q_{11}+Q_{\textrm{TXP}}$ and the enthalpy balance $Q_{12}h_{12}
=Q_{11}h_{11}+Q_{\textrm{TXP}}h_{\textrm{TXP}}$, where we use CoolProp to determine the respective specific
enthalpies.
\section{Joule-Thomson valve\break and liquid extraction}
\label{sec:JT}
Throttling the gas through a JT-valve between point~8 and~9 in Fig.~\ref{fig:FL} causes
the pressure to drop from $P_{8}$ to $P_{9}$, which causes the temperature to drop in such 
a way that the enthalpy in the process remains constant. We model this process using 
CoolProp with the following code snippet
\begin{verbatim} 
  h8=PropSI('H','P',P8,'T',T8,'Helium');
  h9=h8;
  T9=PropSI('T','P',P9,'H',h9,'Helium');
\end{verbatim}
Here {\tt PropSI()} is the MATLAB interface to the CoolProp library. The function receives
two parameters and returns any other, here we supply the initial pressure $P_8$ and
temperature $T_8$ to return the specific enthalpy $h_8$, which stays constant $h_9=h_8.$
In the second call, we calculate the temperature $T_9$ in the reservoir from the known reservoir pressure
$P_9$ and the specific enthalpy $h_9.$
\par
After using CoolProp to determine the specific enthalpies in the reservoir, $h_{\textrm{liq}}$ for
the liquid and $h_{\textrm{gas}}$ for the gas, we obtain the extracted fraction $y$ from solving the
enthalpy balance $h_9=y h_{\textrm{liq}}+(1-y)h_{\textrm{gas}}$ for $y$ and extract the fraction $y$ of the
mass flow $Q_9$ from the gas $Q_{10}$ that returns through the cold side of the heat
exchangers to the compressor.
\section{Results}
\begin{figure}[htb]
\begin{center}
\includegraphics*[width=0.99\columnwidth]{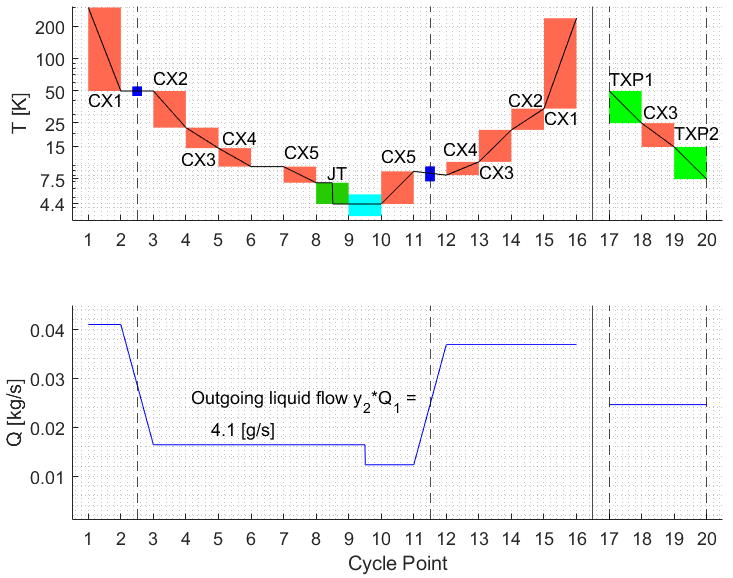}
\end{center}
\caption{\label{fig:RLN}Performance with LN$_2$ pre-cooling.}
\end{figure}
After testing our simulation against textbook examples (Linde-Hampson, Claude, Collins)
and carefully monitoring the balance of the mass flow and enthalpies at all stages, first without liquid nitrogen (LN$_2$) pre-cooling and then with LN$_2$ pre-cooling. The 
result of such a simulation is shown in Fig.~\ref{fig:RLN}. The horizontal axis labels
the points that correspond to those indicated on Fig.~\ref{fig:FL} and on the upper panel
we show the temperature at each point, with the effect of each heat exchanger CXn clearly
indicated by the vertical size of the orange box that signifies the temperature change.
The Joule-Thomson valve is indicated by a green box and the reservoir as a light blue box.
Note also the dark blue dot between point~2 and~3, that shows the branch-off points of
the mass flow through the turbo-expanders, separately shown on the far right of the plot
between points~17 and~20, where point~17 corresponds to the blue dot between points~2
and~3. Point~20 links to the main branch between points~11 and~12, where the gas flows
mix, as described above.
\par
\begin{figure}[htb]
\begin{center}
\includegraphics*[width=0.99\columnwidth]{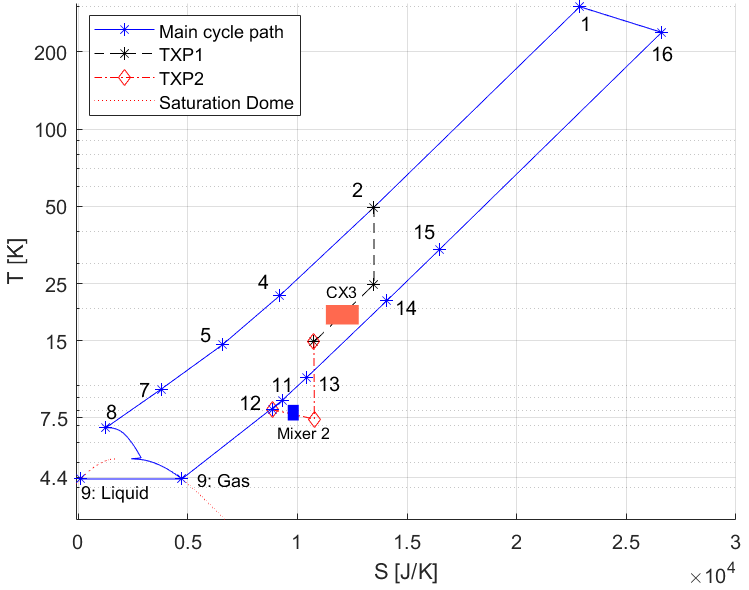}
\end{center}
\caption{\label{fig:TS}TS-diagram with LN$_2$ pre-cooling.}
\end{figure}
We also perform checks such that the enthalpy and gas flow are conserved over the simulated cycle. We then use CoolProp to calculate the entropy $S$ at each point in the cycle and show
the temperature plotted versus the entropy in a $TS$-diagram in Fig.~\ref{fig:TS}, where the
numbers correspond to those in Figs.~\ref{fig:FL} and~\ref{fig:RLN}. This type of diagram is useful to visualize all the heat transfers in the whole cycle. Helium gas in the return flow from the cold box exits at point 16 close to ambient temperature and is heated up further in the piping back to the compressor at point 1. Point 1 to 8 and 9 to 16 respectively represent the isobaric cooling and heating of the gas through the heat exchangers. The separate turbo-expander flow is also visible. Point 8 to 9 describe the isenthalpic process in the JT valve, until
the gas reaches the saturation dome (marked in Fig.~\ref{fig:TS}) and follows the saturated vapour line, whereas some of the gas is left as liquid in the reservoir. 
\par
We adjust the model parameter values, such as the heat exchange design parameter $C'$, to match the performance of the real liquefier. The simulated value of the yield over the whole cycle is found to be 6.1\% without LN$_2$ pre-cooling and 10\% with LN$_2$ pre-cooling, with compressor pressure $P_1 = 12$ bar(a) and reservoir pressure $P_9 = 1.12$ bar(a). The respective values for the yield in the real liquefier is 5-6\% and 10\%. Although temperature and gas sensors are not installed at all cycle points, the simulations show similar values where there are such sensors. 
\par

\par 
In order to find the theoretical maximum yield of the cycle, we vary the model parameters in different combinations. Changing the pressures $P_1$ and $P_9$ as well as the initial gas flow $Q_1$ down to 50\% of their initial values has little impact on the yield $y$ in the simulations, a few percentage points at most. The maximum simulated yield of 17.6\% is obtained by increasing the coupling $C_1'$ (between the hot flow and the cold return flow) in CX1, decreasing the coupling $C_2'$ with the LN$_2$ cooling in CX1, and slightly increasing the turbo-expander mass flow fraction $x$. Although the real heat exchangers have technical limitations in how much $C'$ can be increased, these optimizations can indicate possible performance improvements for the liquefaction. 
\par 

  
\section{Conclusions}
We developed a theoretical model of the helium liquefier in the FREIA Laboratory in MATLAB, starting from enthalpy conservation. The main objective was to find the unknown parameters not specified in the manual of the manufacturer. All the components in the thermodynamic cycle were represented by matrices, except the Joule-Thomson valve for which we used the CoolProp library for the non-linear mapping of the state variables. We observed simulated liquefaction yields similar to the real liquefier, with and without LN$_2$ pre-cooling. Adjusting model parameters allowed us to obtain higher yields and might indicate the promising points to improve the performance of the liquefier.  
\bibliographystyle{plain}

\begin{thebibliography}{M}
%
\bibitem{FREIA}
  R. Ruber et al., {\em Accelerator Development at the FREIA Laboratory},
  \url{https://arxiv.org/abs/2103.05265}
\bibitem{SPOKE}
  H. Li, et al., {\em RF Performance of the spoke prototype cryomodule
    for ESS,} FREIA Report 2019/08;
  \url{http://uu.diva-portal.org/smash/record.jsf?pid=diva2%3A1427442}.
\bibitem{CC}
  A. Miyazaki et al., {\em First cold test of a crab cavity at the GERSEMI
    cryostat for the HL-LHC project,}
  \url{http://uu.diva-portal.org/smash/record.jsf?pid=diva2%3A1501277}
\bibitem{COR}
    K. Pepitone, in preparation.
\bibitem{EW}
  E. Waagaard, {\em Benchmarking a cryogenic code for the FREIA helium liquefier,}
  FREIA Report 2020/01; available from
  \url{http://uu.diva-portal.org/smash/record.jsf?pid=diva2%3A1438754&dswid=4228}.   
\bibitem{CP}
I. Bell et al., {\em Pure and Pseudo-pure Fluid Thermophysical Property Evaluation and
  the Open-Source Thermophysical Property Library CoolProp,} Ind. Eng. Chem. Res. {\bf 52}
(2014) 2498; and at \url{http://www.coolprop.org}
\bibitem{FLYNN}
  T. Flynn, {\em Cryogenic engineering,} CRC Press, Baton Roca, Fl., 2004.
%
\end{thebibliography}

%
%
\end{document}